\documentclass[twocolumn,showpacs,preprintnumbers,amsmath,amssymb]{revtex4}%\begin{widetext}

\newcommand{\bea}{\begin{eqnarray}}
\newcommand{\eea}{\end{eqnarray}}
\usepackage{epsfig}
\usepackage{graphicx}% Include figure files
\usepackage{dcolumn}% Align table columns on decimal point
\usepackage{bm}% bold math
\let\jnfont=\rm
\def\NPB#1,{{\jnfont Nucl.\ Phys.\ B }{\bf #1},}
\def\PLB#1,{{\jnfont Phys.\ Lett.\ B }{\bf #1},}
\def\EPJC#1,{{\jnfont Eur.\ Phys.\ Jour.\ C }{\bf #1},}
\def\PRD#1,{{\jnfont Phys.\ Rev.\ D }{\bf #1},}
\def\PRL#1,{{\jnfont Phys.\ Rev.\ Lett.\ }{\bf #1},}
\def\MPLA#1,{{\jnfont Mod.\ Phys.\ Lett.\ A }{\bf #1},}
\def\JPG#1,{{\jnfont J.\ Phys.\ G}{\bf #1},}
\def\CTP#1,{{\jnfont Commun.\ Theor.\ Phys.\ }{\bf #1},}
\def\JHEP#1,{{\jnfont JHEP \ }{\bf #1},}
\def\NPPS#1,{{\jnfont Nucl.\ Phys.\ Proc.\ Suppl.\ }{\bf #1},}

\def\lsim{\mathrel{\mathpalette\oversim<}}
\def\gsim{\mathrel{\mathpalette\oversim>}}
\def\oversim#1#2{\lower0.5ex\vbox{\baselineskip0pt\lineskip0pt
  \lineskiplimit0pt\everycr{}\tabskip0pt
  \halign{$\mathsurround0pt #1\hfil##\hfil$\crcr #2\crcr\sim\crcr}}}
\begin{document}

\preprint{\parbox{1.2in}{\noindent  arXiv:0901.3818}}

\title{\ \\[1mm] Residual effects of heavy sparticles in the bottom quark Yukawa coupling:\\
                  a comparative study for the MSSM and NMSSM}

\author{\ \\[1mm] Wenyu Wang$^1$, Zhaohua Xiong$^1$,
                  Jin Min Yang$^2$ \\ ~}

\affiliation{
$^1$ Institute of Theoretical Physics, College of Applied Science,
              Beijing University of Technology, Beijing 100020, China\\
$^2$  Key Laboratory of Frontiers in Theoretical Physics,
      Institute of Theoretical Physics, Chinese Academy of Sciences,
              Beijing 100190, China}

\begin{abstract}
If the sparticles are relatively heavy (a few TeV) while the Higgs
sector is not so heavy ($m_A$ is not so large), the Higgs boson
Yukawa couplings can harbor sizable quantum effects of sparticles
and these large residual effects may play a special role in
probing supersymmetry at foreseeable colliders. In this work,
focusing on the supersymmetric QCD effects in the $hb\bar b$
coupling  ($h$ is the lightest CP-even Higgs boson), we give a
comparative study for the two popular
supersymmetric models: the MSSM and NMSSM. While for both models the
supersymmetric QCD can leave over large residual quantum effects
in $hb\bar b$ coupling, the NMSSM can allow for a much broader
region of such effects. Since these residual effects can be over
$20\%$ for the $hb\bar b$ coupling (and thus over $40\%$ for the
ratio $Br(h\to b\bar b)/Br(h\to \tau^+ \tau^-)$), future
measurements may unravel the effects of heavy sparticles or even
distinguish the two models.

\end{abstract}
\pacs{14.80.Cp,12.60.Fr,11.30.Qc}

\maketitle

%%%%%%%%%%%%%%%%%%%%%%%%%%%%%%%%%%%%%%%%%%%%%%%%%%%%%%%%%%%%%%%%%%%%%
\section{Introduction}

Supersymmetry is a prime candidate for new physics beyond the Standard
Model (SM).
Among various supersymmetric models the most extensively studied is
the minimal supersymmetric model (MSSM)  \cite{MSSM}.
Another popular supersymmetric model, which may be equally or more
attractive compared with the MSSM, is the next-to-minimal supersymmetric
model (NMSSM) \cite{NMSSM} since it can solve the $\mu$-problem
and alleviate the little hierarchy. These models will soon be put to
the test at the LHC. Thus, their phenomenological study is important and urgent.

Although the most convincing evidence of supersymmetry is the
detection of sparticle (sfermions, gauginos or Higgsinos)
productions, the indirect probe through detecting the quantum
effects of virtual sparticles in some measurable interactions will
play a complementary role. If the sparticles are relatively heavy
(say above a few TeV) and hence cannot be directly detected at the
LHC, the indirect probe through quantum effects could be
important. For this end, the Higgs boson Yukawa interactions may
play a special role since they can harbor sizable quantum effects
of heavy sparticles when the Higgs sector is not so heavy (i.e.
$m_A$ is not so large). The dominant quantum effects of sparticles
are from the supersymmetric QCD interaction and in the literature
the calculations have been performed in the MSSM for such
supersymmetric QCD effects in the Higgs boson Yukawa couplings
\cite{hbb-htb,hbb-nondec} and the associated Higgs production
processes at the LHC \cite{hbb-process}. The studies in the
decoupling limit with heavy sparticles showed that for a light
Higgs sector  ($m_A$ is not large) the supersymmetric QCD can
leave over large residual quantum effects in both the decay $h\to
b\bar b$ \cite{hbb-nondec} and the productions at the LHC
\cite{hbb-process}. Given the popularity of the NMSSM, it is
necessary to extend the study to the NMSSM. This is the aim of
this work.

In this work we will focus on the supersymmetric
QCD effects in the $hb\bar b$ coupling ($h$ is the lightest CP-even neutral
Higgs boson) and perform a comparative study for the two popular supersymmetric
models: the MSSM and NMSSM. Such a study is interesting for two points:
\begin{itemize}
\item[(i)] The study of the NMSSM can accommodate the study of the
           MSSM, and
           in some limit the NMSSM results can reduce to the MSSM results.
           One can envisage that the supersymmetric residual effects in the MSSM
           can be magnified in the NMSSM. So the future measurements of such
           effects may be useful in telling the difference of the two models.
\item[(ii)] Compared with the MSSM, the NMSSM may predict a different tree-level
            $hb\bar b$ coupling and different loop contribution.
            In the NMSSM the mass matrix and mixings of the Higgs bosons
            are enriched and thus the components of the lightest CP-even Higgs
            boson $h$ are different from the MSSM values. The residual
            supersymmetric QCD effects in the NMSSM may be larger and
            more interesting.
\end{itemize}
Note that supersymmetry is a decoupling theory
and all low-energy observables will recover their corresponding SM predictions
when the mass scale of all supersymmetric particles (including the masses of
sparticles and $m_A$) take their heavy limits. The large residual quantum
effects of sparticles in the Higgs Yukawa couplings happen only in case
that sparticles are heavy but $m_A$ is light. If both sparticles and $m_A$
take their heavy limit, the residual effects of supersymmetry do vanish.
Since so far such a split scenario (with light Higgs bosons and relatively
heavy sparticles) remains possible, we should check its phenomenological
consequence.

\section{Calculations}
We start our analysis by recapitulating the basics of the NMSSM.
In the NMSSM a singlet Higgs superfield $\hat{S}$ is introduced
and the Higgs terms in the superpotential are given by
\begin{eqnarray}\label{superpotential}
\lambda\hat{S}\hat{H_d}\cdot\hat{H_u}-\frac{\kappa}{3}\hat{S}^3 \, ,
\end{eqnarray}
where $\hat{H}_u$ and $\hat{H}_d$ are the Higgs doublet superfields,
and $\lambda$ and $\kappa$ are the dimensionless constants.
Note that there is no explicit $\mu$-term and an effective $\mu$-parameter
is generated when the scalar component ($S$) of $\hat{S}$ develops a vev $s$:
$ \mu_{eff}= \lambda s$.
The corresponding soft SUSY breaking terms are given by
\begin{eqnarray}\label{soft-term}
-A_\lambda \lambda S H_d\cdot H_u-\frac{A_\kappa}{3}\kappa S^3 +h.c.\, .
\end{eqnarray}
So the scalar Higgs potential is given by
\small
\begin{eqnarray}\label{potential}
V_F &=& |\lambda H_d\cdot H_u- \kappa S^2|^2
     + |\lambda S|^2 \left(|H_d|^2+|H_u|^2 \right)\, , \\
V_D &=&\frac{g_{2}^2}{2} \left( |H_d|^2|H_u|^2-|H_d \cdot H_u|^2 \right) \nonumber\\
    && +\frac{g_1^2+g_2^2}{8} \left( |H_d|^2-|H_u|^2\right)^2\, , \\
V_{\rm soft}&=&m_{d}^{2}|H_d|^2 + m_{u}^{2}|H_u|^2
            + m_s^{2}|S|^2 \nonumber\\
    && - \left( A_\lambda \lambda S H_d\cdot H_u
           + \frac{\kappa}{3} A_{\kappa} S^3 + h.c. \right)\, ,
\end{eqnarray}
\normalsize
where $g_1$ and $g_2$ are the coupling constant
of $U_Y(1)$ and $SU_L(2)$, respectively.
So we can see that in the limit of vanishing $\lambda,~\kappa,~A_\kappa$
and with the input of the effective $\mu_{eff}$, the NMSSM can reduce to
the MSSM.

With the vevs $v_u$, $v_d$ and $s$, the scalar fields are expanded as
\begin{eqnarray}
H_d & =& \left ( \begin{array}{c}
              v_d + \phi_d + i \varphi_d \\
             H_d^- \end{array} \right) \, , \\
H_u & =& \left ( \begin{array}{c} H_u^+ \\
                v_u + \phi_u + i \varphi_u
        \end{array} \right)  \, , \\
S & =& s + \sigma + i \xi \, .
\end{eqnarray}
The mass eigenstates can be obtained by unitary rotations of interaction
states, e.g., for the CP-even neutral mass eigenstates ($h,H_1,H_2$) we
have
\begin{eqnarray}
  \label{eq:evenhiggs}
  \left(h,~H_1,~H_2\right)^T
=\sqrt{2} U^H \left(\phi_d,~\phi_u,~\sigma\right)^T .
\end{eqnarray}
So the lightest CP-even neutral Higgs boson $h$ is composed by
\begin{eqnarray}
h=U^H_{11}\sqrt{2}\phi_d+U^H_{12}\sqrt{2}\phi_u+U^H_{13}\sqrt{2}\sigma.
\label{eq13}
\end{eqnarray}
In the MSSM, $h$ can be decomposed in the same way except
that without the singlet component, $U^H$ is a $2\times 2$ matrix
and can be parameterized in terms of a mixing angle $\alpha$.

The sbottom squared-mass matrix in the NMSSM is same as in
the MSSM with $\mu$ replaced by $\mu_{eff}$. In the basis
of ($\tilde b_L$,$\tilde b_R$), the squared-mass matrix is
given by
\begin{equation}
 \mathcal{M}^2_{\tilde b} = \left(\begin{array}{cc}
    M_L^2 & m_b X_b \\ m_b X_b & M_R^2 \end{array} \right)\, ,
\label{eq:sbottommatrix}
\end{equation}
where
\begin{eqnarray}
    X_b &=& A_b + \mu_{eff} \tan\beta\,, \nonumber \\
    M_L^2 &=& M_{\tilde Q}^2 + m_b^2
    + M_Z^2 (I_3^b - Q_b s^2_W) \cos 2\beta\,,  \nonumber \\
    M_R^2 &=& M_{\tilde b_R}^2 + m_b^2 + M_Z^2 Q_b s^2_W \cos 2\beta\,.
    \label{eq:sbottomparams}
\end{eqnarray}
with $s_W\equiv \sin\theta_W$, $\tan\beta=v_u/v_d$,
$I_3^b$ and $Q_b$ being respectively
the isospin and electric charge of the $b$-quark, $M_{\tilde Q}$
and $M_{\tilde b_R}$ being the soft breaking masses, and $A_b$ being
the soft breaking trilinear coupling.
The sbottom mass eigenstates $(\tilde b_1,~\tilde b_2)$
are obtained by the unitary rotation of the interaction eigenstates:
\begin{eqnarray}
\left( \begin{array}{l} \tilde b_L \\ \tilde b_R \end{array} \right)
 =Z_D\left( \begin{array}{l} \tilde b_1 \\ \tilde b_2 \end{array} \right)
 = \left( \begin{array}{cc}
    \cos \theta_{\tilde b} & -\sin \theta_{\tilde b} \\
    \sin \theta_{\tilde b} & \cos \theta_{\tilde b}
  \end{array} \right)
 \left( \begin{array}{l} \tilde b_1 \\ \tilde b_2 \end{array} \right)
\end{eqnarray}
where the unitary matrix $Z_D$ parameterized by a mixing angle
$\theta_{\tilde b}$ diagonalizes the mass matrix in
Eq.(\ref{eq:sbottommatrix}).

\begin{figure}[htbp]
%\scalebox{1}{
\epsfig{file=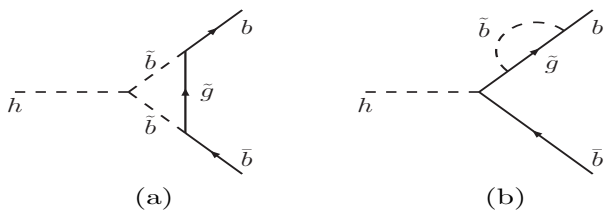,height=3cm,width=8.5cm}
\vspace{-3mm}
\caption{Feynman diagrams for the SUSY QCD
        corrections to the vertex $hb\bar b$.}
\label{eff_hbb}
\end{figure}

The coupling of sbottoms with the singlet component $\sigma$ of the
Higgs boson $h$ comes from the F-term of the superpotential, given by
\begin{eqnarray} \label{eq:sqq}
-\frac{g_2 m_b}{2 m_W \cos\beta}
\lambda v_u\sqrt{2}\sigma\tilde b^\ast_L\tilde b_R+ h.c.\, .
\end{eqnarray}
In terms of the mass eigenstates of sbottoms, the vertex $h\tilde b_i\tilde b_j $
takes the form
\small
\begin{eqnarray} \label{eq:hsqsq}
V_{h\tilde b_i\tilde b_j} &=&
-i\frac{g_2 m_b}{2 m_W \cos\beta}
\left(U^H_{11}c_d^{ij}+U^H_{12}c_u^{ij}+U^H_{13}c_s^{ij}\right)
\end{eqnarray}
\normalsize
where $c_s^{ij}$ is only for the NMSSM (or equivalently set
$U^H_{13}=0$ for the MSSM), while $c_{d}^{ij}$ and $c_{u}^{ij}$ are
present for both models. They are given by
\begin{eqnarray} \label{eq:cd}
c_d^{ij} &=& \left[2m_b-\frac{m^2_Z \cos^2\beta}{m_b}(1
-\frac{2}{3}s_W^2)\right]Z_D^{1i}Z_D^{1j} \nonumber\\
&& +\left[2m_b-\frac{2m^2_Zs_W^2\cos^2\beta}{3m_b}\right]Z_D^{2i}Z_D^{2j}\nonumber\\
&& +A_b\left(Z_D^{1i}Z_D^{2j}+Z_D^{2i}Z_D^{1j}\right) \, ,\\
c_u^{ij} &=& \frac{m^2_Z\sin 2\beta}{2m_b}\left(1
-\frac{2}{3}s_W^2\right)Z_D^{1i}Z_D^{1j}   \nonumber \\
&& +\frac{m^2_Zs^2_W\sin 2\beta}{3m_b}Z_D^{2i}Z_D^{2j} \nonumber \\
&& +\mu_{eff}\left(Z_D^{1i}Z_D^{2j}+Z_D^{2i}Z_D^{1j}\right) \, , \label{eq:cu} \\
c_s^{ij}&=&\frac{\sqrt{2}\lambda m_W \sin\beta}{g_2}
\left(Z_D^{1i}Z_D^{2j}+Z_D^{2i}Z_D^{1j}\right). \label{eq:css}
\end{eqnarray}
Now we calculate the SUSY QCD corrections to the
vertex $hb\bar b$. At tree level it takes the same
form in both the MSSM and NMSSM, given by
\begin{eqnarray} \label{eq:treehbb}
 V^0_{hb \bar b}=-i \frac{g_2 m_bU^H_{11}}{2 m_W \cos\beta}
\end{eqnarray}
The one-loop SUSY QCD corrections come from the Feynman diagrams shown in
Fig.\ref{eff_hbb} where we do not show the self-energy loop of $\bar b$.
%%%Fig.1%%%%%%%%%%%%%%%%%%%%%%%%%%%%%%%%%%%%%%%%%%%%%%%%%%%%%%%
%%%%%%%%%%%%%%%%%%%%%%%%%%%%%%%%%%%%%%%%%%%%%%%%%%%%%%%%%%%%%%%
In our calculations we use the on-shell renormalization scheme
\cite{denner} and take the external $b$ and $\bar b$ quarks on
shell to get the effective vertex. With the corrections the
effective vertex takes the form
\begin{eqnarray} \label{eq:eff_hbb}
V_{h b \bar{b}}= V^0_{h b \bar{b}} \left(1+\Delta^{\rm v} +\Delta^{\rm ct}\right) ,
\end{eqnarray}
where $\Delta^{\rm v}$ denotes the vertex correction from Fig.\ref{eff_hbb}(a)
and $\Delta^{\rm ct}$ is the counter term from the renormalization of
$m_b$ and the wave functions of $b$ and $\bar b$.
They are given by
\small
\begin{eqnarray}
\Delta^{{\rm v}}
  &=& \frac{\alpha_s}{3\pi} \left(c_d^{ij}+\frac{U^H_{12}}{U^H_{11}}c_u^{ij}
      +\frac{U^H_{13}}{U^H_{11}}c_s^{ij}\right)\nonumber \\
  && \times
    \left[m_b\left( Z_D^{1i}Z_D^{1j}+Z_D^{2i}Z_D^{2j}\right)C_{11}
      +m_{\tilde g}\left(Z_D^{1i}Z_D^{2j} \right.\right.\nonumber \\
  && \left.\left.+Z_D^{2i}Z_D^{1j}\right)C_0\right]
     (m^2_b, m^2_h,m^2_b,m^2_{\tilde g},m^2_{\tilde b_{i}},m^2_{\tilde b_{j}}), \label{eq24} \\
\Delta^{{\rm ct}}&=& - \frac{\alpha_s}{3\pi}
    \Biggl\{ \frac{m_{\tilde g}}{m_b} \sin(2\theta_{\tilde b})
    \big[B_0 (m_b^2,m_{\tilde g}^2,m_{\tilde b_1}^2) \nonumber \\
 && - B_0(m_b^2,m_{\tilde g}^2,m_{\tilde b_2}^2) \big]
    - 2 m_b^2 \big[B_1^{\prime}(m_b^2,m_{\tilde g}^2,m_{\tilde b_1}^2) \nonumber \\
 && + B_1^{\prime}(m_b^2,m_{\tilde g}^2,m_{\tilde b_2}^2) \big]
 - 2 m_b m_{\tilde g} \sin(2\theta_{\tilde b})\nonumber \\
 && \times
    \big[B_0^{\prime}(m_b^2,m_{\tilde g}^2,m_{\tilde b_1}^2)
  - B_0^{\prime}(m_b^2,m_{\tilde g}^2,m_{\tilde b_2}^2) \big] \Biggr\} ,
    \label{eq:con}
\end{eqnarray}
\normalsize
where $\alpha_s$ is the strong coupling constant,
$m_{\tilde g}$ is the gluino mass, and $B_{0,1}$, $B'_{0,1}$ and
$C_{0,11}$ are the scalar loop functions \cite{Hooft} which can be
calculated by using LoopTools \cite{looptools}.

\section{Numerical results}
In our calculation we use the package NMSSMTools \cite{nmssmtools}
for the mass spectrum and the rotation matrix of the Higgs fields
(to get the corresponding MSSM results we take the limit of very
small values for $\lambda$, $\kappa$ and $A_\kappa$, and results
are checked by using the package FeynHiggs \cite{feynhiggs}).
To study the decoupling limit of SUSY particles, we assume
all the soft breaking mass parameters
($M_{\tilde Q}$,$M_{\tilde b_R}$,$m_{\tilde g}$,$A_b$) and the
parameter $\mu_{eff}$ are degenerate, which are  collectively denoted
by $M_{SUSY}$.
Then in the MSSM the SUSY parameters are ($M_{SUSY}$, $M_A$, $\tan\beta$),
while in the NMSSM there are three additional parameters
($\lambda$,~$\kappa$,~$A_\kappa$). In our calculations we
scan over these three additional parameters in the ranges
\begin{eqnarray} \label{eq:lkak}
-0.5<\lambda, \kappa<0.5, ~-500 {\rm ~GeV}<A_\kappa<500 {\rm ~GeV}
\end{eqnarray}
Note here we did not consider a large $\lambda$ or $\kappa$.
Theoretically, the requirement of perturbativity
up to some cut-off scale will set upper bounds
on $\lambda$ and $\kappa$ at weak scale (if the cut-off scale
is chosen to be the GUT scale, a stringent bound
$\lambda^2+\kappa^2 \lsim 0.5$ is obtained \cite{Miller:2003ay}).
Phenomenologically, a large $\lambda$ or $\kappa$ will incur
stringent constraints from current experiments (see the last
reference in \cite{NMSSM}).

\begin{figure}[htbp]
\scalebox{0.5}{\epsfig{file=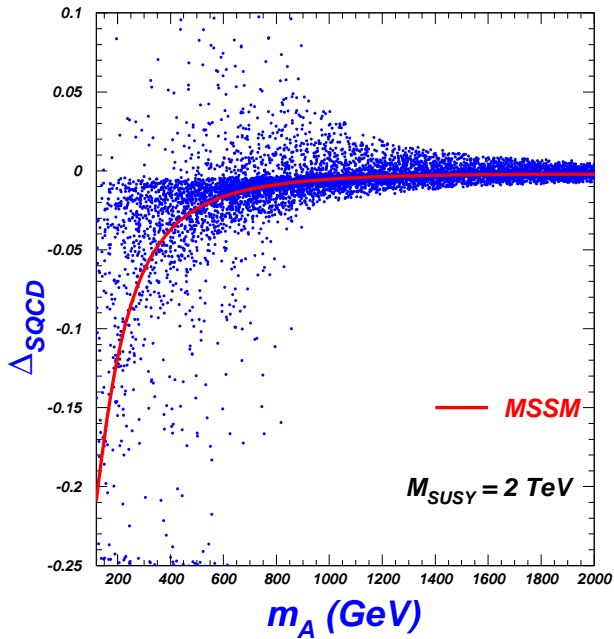}}
\vspace{-0.5cm}
\caption{SUSY QCD corrections to $hb\bar b$ coupling versus $m_A$
         in the MSSM and NMSSM for $\tan\beta=20$.
         The scatter plots are for the NMSSM while the curves for the MSSM.}
\label{dcp_as}
\end{figure}

\begin{figure}[htbp]
\hspace{-0.5cm}
\vspace{-0.4cm}
\scalebox{0.47}{\epsfig{file=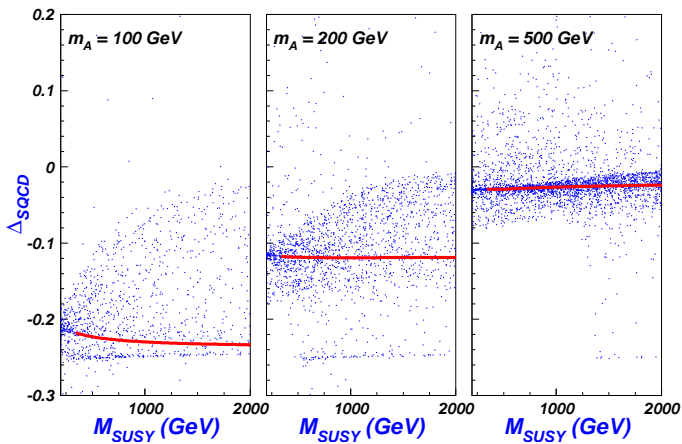}}
\caption{Same as Fig. \ref{dcp_as}, but versus $M_{SUSY}$ for fixed values
         of $m_A$.}
\label{d_a}
\end{figure}

Before displaying the numerical results, we make some clarifications
regarding to our numerical calculations:
\begin{itemize}
\item[(1)] In our calculations we used the package NMSSMTools \cite{nmssmtools}
           which considered the loop corrections (especially the stop/sbottom loops)
           to the effective potential, the masses and mixing angles of the Higgs bosons.
           These corrections are important and cannot be ignored.
\item[(2)] Since we used the package NMSSMTools in which the Higgs mass matrices
           are diagonalized numerically, we did not use any approximate
           unitary transformation for our calculation.
\item[(3)] For the b-quark mass $m_b$, in the package NMSSMTools (and thus in our
           calculation), it is taken as the running mass $m_b (Q)$ (we take $Q=M_{SUSY}$),
           which means that the sizable QCD loop effects are taken into account.
           Actually, although both the tree-level coupling $V^0_{hb\bar{b}}$
           and the one-loop SUSY QCD contributions $\delta V_{hb\bar{b}}$
           are proportional to $m_b$ and thus very sensitive to the value
           of  $m_b$, the relative correction effects $\delta V_{hb\bar{b}}/V^0_{hb\bar{b}}$
           (displayed in our numerical results) are not so sensitive to the
           value of $m_b$.
\end{itemize}
Now we present some numerical results.
To see the general feature of the corrections,
we first switch off the experimental constraints on the parameter space
and in Figs.\ref{dcp_as} and \ref{d_a} display the resluts of
the relative correction effects
\begin{eqnarray}
\Delta_{SQCD}\equiv \frac{\delta V_{hb\bar{b}}}{V^0_{hb\bar{b}}}
   = \Delta^{\rm v}+\Delta^{\rm ct}.
\label{eq27}
\end{eqnarray}
We see that for a light $m_A$ the SUSY QCD with a large $M_{SUSY}$
can leave over sizable effects in the coupling  $hb\bar b$. But as
$m_A$ gets heavy, such residual effects of SUSY QCD become small
in magnitude, showing the decoupling behavior of supersymmetry.
Compared with the MSSM results, the effects in the NMSSM can vary
in a much broader region. While the corrections are always
negative in the MSSM, in the NMSSM the corrections can be both
negative and positive.
%%%%%%Fig.2%%%%%%%%%%%%%%%%%%%%%%%%%%%%%%%%%%%%
%%%%%%%%%%%%%%%%%%%%%%%%%%%%%%%%%%%%%%%%%%%%%%%%%%%
%\begin{widetext}
%%%%%%Fig.3%%%%%%%%%%%%%%%%%%%%%%%%%%%%%%%%%%%%
%%%%%%%%%%%%%%%%%%%%%%%%%%%%%%%%%%%%%%%%%%%%%%%%%%%%%%%%%%%%
\begin{table}[tb]
\caption{Some sample points in the parameter space of the NMSSM,
in which the SUSY QCD loop effects $\Delta_{SQCD}$ can be of very different size.
The mass parameters are in unit of GeV.
 Here we fixed $M_A=200$ GeV and  $\tan\beta=20$.
With such fixed parameters and $M_{SUSY}$ varying from 400 GeV to 2 TeV,
the value of $m_h$ in the MSSM varies from 106 GeV to 119 GeV while
the corresponding values of $\Delta_{SQCD}$ in the MSSM are about $-11\%$.}
\begin{tabular}{cccccccc}
\hline\hline
  \      &\        &     &      &\       &   &   &   \\
  $\lambda$&$\kappa$&$A_\kappa$&$M_{SUSY}$&$U^H_{12}/U^H_{11}$&
$U^H_{13}/U^H_{11}$&$m_h$ &$\Delta_{SQCD}$\\
   \      &\        &     &      &\       &   &   &  (\%)\\
\hline
    0.106& -0.215~ &  446   &   1788 &16.2    &-0.023~ & 117.4    &  ~-4.1\\
   -0.116~~&  0.180 &  333   &   1575 &14.7    &0.042  & 116.3    &  ~-6.0\\
    0.133& -0.256~ &  393   &   1145 &14.8    &-0.046~ & 114.7    &  ~-6.1\\
    0.123& -0.164~ &  460   &   1568 &14.1    &-0.062~ & 116.1    &  ~-6.8\\
   -0.103~~& -0.072~ & -294~   &   1544 &10.1    &0.118  & 114.8    &  -12.0\\
    0.107&  0.073 & -175~   &   1831 &9.7    &-0.100~ & 115.2   &  -12.4\\
    0.126&  0.101 & -426~   &   1738 &7.3    &-0.069~ & 114.8    &  -15.5\\
   -0.106~~& -0.123~ & -377~   &   1401 &5.4    &0.024  & 115.3    &  -18.0\\
    0.110&  0.130 & -324~   &   1419 &4.1    &-0.017~ & 114.9    &  -19.7\\
    0.106&  0.406 & -304~   &   ~623  &0.5    &0.002  & ~99.4    &  -24.5\\
\hline \hline
\end{tabular}\label{table-1}
\end{table}
%%%%%%%%%%%%%%%%%%%%%%%%%%%

In order to figure out in what areas of the NMSSM parameter space
the one-loop SUSY QCD effects on the $hb\bar{b}$ coupling are
sizable, we present a set of sample points in Table \ref{table-1}.
We see that for the SUSY QCD loop effects on the $hb\bar{b}$
coupling to be sizable in the NMSSM, the ratio $U_{12}^H/U_{11}^H$
( $U_{12}^H$ and $U_{11}^H$ are respectively the components of
$\phi_u$ and $\phi_d$ in $h$, as defined in Eq.\ref{eq13}) plays
the key role for the following reasons. As shown in
Eq.(\ref{eq27}), $\Delta_{SQCD}$ is composed of two parts: the
counter-term part $\Delta^{\rm ct}$ and the vertex-loop part
$\Delta^{\rm v}$. We found that in most of the parameter space
allowed by the LEP constraints, $\Delta^{\rm ct}$ is negative and
dominant in size, which is independent of how $h$ is composed of.
Whereas,  $\Delta^{\rm v}$ is positive and cancel $\Delta^{\rm
ct}$ to some extent. The size of  $\Delta^{\rm v}$ can be enhanced
by the ratio $U_{12}^H/U_{11}^H$, as shown in  Eq.(\ref{eq24}).
Although the ratio $U_{13}^H/U_{11}^H$ can also enhance the size
of  $\Delta^{\rm v}$, its effect is suppressed by the smallness of
$U_{13}^H$ and $\lambda$ in $c_s^{ij}$ (in our scan we found that
both $\lambda$ and $U_{13}^H$ are small in order to satisfy the
LEP constraints encoded in the  NMSSMTools). Therefore, as
$U_{12}^H/U_{11}^H$ gets large,
 $\Delta^{\rm v}$ becomes more sizable and, due to its cancellation effect,
the total correction effects become less sizable.
Note that although the value of $m_h$ is below 114 GeV for some points,
these points can still satisfy the LEP constraints encoded in the NMSSMTools
because the LEP bound on the MSSM $m_h$ is 92 GeV (see Fig.2 in
the paper by Barger et. al. in Ref.\cite{NMSSM}).  By the way,
some comprehensive studies on the phenomenology
of the NMSSM Higgs sector (which may be quite different from the
phenomenology of the MSSM Higgs sector) have been performed in Ref.\cite{NMSSM},
where all the Higgs couplings and all the decay modes as well as productions
have been intensively studied.

From Table \ref{table-1} we see that although $\lambda$ is scanned
in the range of $-0.5\sim 0.5$, only small values of $\lambda$
($\sim 0.1$ in magnitude) survived the LEP constraints encoded in
the NMSSMTools. As $\lambda$ gets large, the mixing between
singlet and doublet Higgs fields becomes large (although the
mixing is not totally determined by $\lambda$ and other parameters
are also contributing) and, consequently, is more constrained by
LEP experiments \cite{NMSSM}. Such a small  $\lambda$, together
with other NMSSM parameters appearing in the mass matrix of the
Higgs fields, leads to a small mixing  between singlet and doublet
Higgs fields. As shown in Table \ref{table-1}, the singlet
component $U_{13}^H$ in $h$ is small, which means that the
lightest Higgs boson $h$ is not the singlet-like one and instead
it is doublet-dominant (MSSM-like). For the $hb\bar{b}$ coupling
with such a MSSM-like $h$, the NMSSM can still allow for quite
different SUSY QCD effects compared with the MSSM predictions. The
reason is that, as discussed above, the SUSY QCD effects are
sensitive to the ratio $U_{12}^H/U_{11}^H$, which is different in
these two models (even for a small $\lambda$, the NMSSM can still
allow for different-size components $U_{12}^H$ and $U_{11}^H$ in
$h$ because other NMSSM parameters are also contributing to the
mass matrix and mixings of the Higgs fields).

\begin{figure}[htbp]
\scalebox{0.5}{\epsfig{file=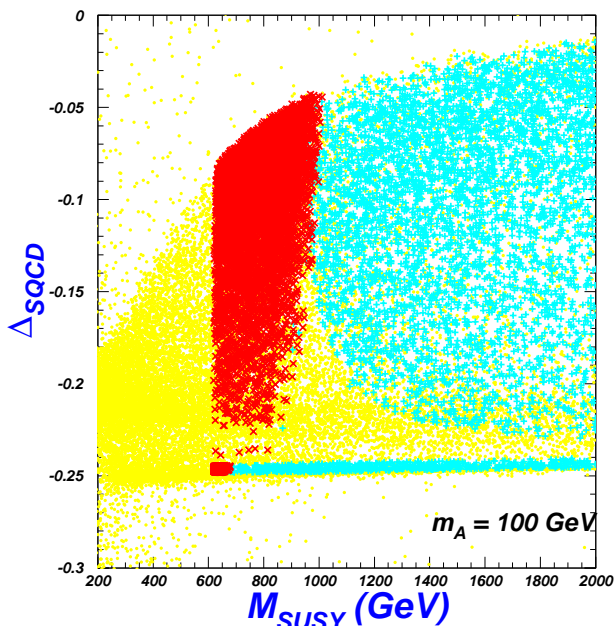}}
\vspace{-0.5cm}
\caption{Same as Fig. \ref{d_a}, but for $m_A=100$ GeV in the NMSSM.
         The dark-shaded region denoted by '$\times$' (red in color)
         are allowed
         by the experimental constraints considered in
         the NMSSMTools \cite{nmssmtools}.
         The light-shaded region denoted by '$\bullet$' (yellow in color)
         are excluded by
         LEP experiment or B physics,
         and the medium-shaded region denoted by '$+$'
         (sky-blue in color)
         are excluded by the muon $g-2$.}
\label{con_as}
\end{figure}

When we switch on the comprehensive experimental constraints encoded in
the NMSSMTools \cite{nmssmtools},
we obtain the results of $\Delta_{SQCD}$ displayed in Fig.\ref{con_as}.
These experimental constraints are comprehensive, including the LEP II
searches for the Higgs boson, various B-decays and the muon anomalous
magnetic moment (muon $g-2$).
%%%%%%%Fig.4 %%%%%%%%%%%%%%%%%%%%%%%%%%%%%%%%%%%%%%%%%%%%%%%%%%%%%
 %%%%%%%%%%%%%%%%%%%%%%%%%%%%%%%%%%%%%%%%%%%%%%%%%%%%%

From Fig.\ref{con_as} we see that in the special scenario under
our consideration (all soft breaking mass parameters are
degenerate), $M_{SUSY}$ is constrained in a certain range. The
range of $M_{SUSY}\lsim 600$ GeV is not allowed by LEP
experiments, while $M_{SUSY}\gsim 1$ TeV cannot explain the muon
$g-2$ data (we require the supersymmetric effects to account for
the deviation of $a_\mu^{exp} - a_\mu^{SM} = ( 29.5 \pm 8.8 )
\times 10^{-10} $ at $2 \sigma$ level). In the allowed region of
the parameter space, the SUSY QCD corrections to the $hb\bar b$
coupling can still be significant, over $20\%$ in magnitude. If we
switch off the muon $g-2$ constraint ( the hadronic contribution
to $a_\mu^{SM}$ is not so certain \cite{Miller}), the allowed
parameter space gets much broader, as shown in  Fig.\ref{con_as}.
In our study we did not require supersymmetry to explain various
plausible evidences of dark matter.

Note that the SUSY contributions to the muon $g-2$ are sensitive to
the soft masses in slepton sector and chargino sector, but not dependent
on the squark or gluino mass involved in our SUSY QCD loops.
The stringent constraint from  the muon $g-2$ data shown in  Fig.\ref{con_as}
comes from our simple assumption that all soft masses (in the squark sector,
the slepton sector and the gaugino sector) are degenerate. Of course, the muon $g-2$
constraint is not necessary; without such an assumption of degeneracy, the
constraint is lifted.

Since the SUSY QCD residual effects can be over $20\%$ for the $hb\bar b$
coupling and thus over $40\%$ for the ratio of the branching fractions
(SUSY QCD does not contribute to $h\to \tau^+ \tau^-$ at one-loop level)
\begin{eqnarray}
R_{b/\tau}= \frac{Br(h\to b\bar b)}{Br(h\to \tau^+ \tau^-)}\, ,
\end{eqnarray}
future measurements at the LHC or ILC may unravel such
supersymmetric effects. This ratio $R_{b/\tau}$ is proposed in
\cite{ratio} as a probe to new physics. To measure this ratio at
the LHC, one may count the event numbers of the production of
$hb\bar b$ followed respectively by the decay $h\to b\bar b$ and
$h\to \tau^+ \tau^-$, and the ratio of these two event-numbers can
be a measure of $R_{b/\tau}$ (the difference of efficiency for
$b$-tagging and $\tau$-tagging should be taken into account).

Finally, we make some remarks regarding to our results:
\begin{itemize}
\item[(1)] We only investigated the SUSY QCD corrections, which is
${\cal O}(\alpha_s)$ and thus should be the most important among
the SUSY corrections. Among the SUSY electroweak corrections, the
Higgs-top Yukawa corrections, which are ${\cal O}(\alpha
\frac{m_t^2}{m_W^2})$, may also be sizable although they are
seemingly not as large as the ${\cal O}(\alpha_s)$ SUSY QCD
corrections. With the consideration of the SUSY electroweak
corrections, our conclusion will remain qualatively unchanged.
\item[(2)] Since our main interest is the residual effects of
heavy sparticles, we assumed all soft SUSY-breaking mass
parameters are equal to $M_{SUSY}$. This is a very strong
simplifying assumption. If we lift such an  assumption and
consider the multiple free soft parameters, we will obtain the
results which numerically are different to some extent while
qualitatively exhibit the same feature, i.e., the NMSSM allows
broader corrections than the MSSM, because the NMSSM parameters
can complicate the mass matrix and mixings of Higgs fields.
\item[(3)] We should stress again that supersymmetry is a
decoupling theory and the large residual quantum effects of
sparticles in the Higgs Yukawa couplings are present only in case
of a light $m_A$. If both $M_{SUSY}$ and $m_A$ take their heavy
limit, the SUSY effects will vanish. Such a decoupling behavior is
similar in the NMSSM and MSSM, as shown in Fig.2.
\end{itemize}

\vspace*{0.5cm}
\section{Summary}
We focused on the SUSY QCD effects in the $hb\bar b$ coupling and
performed a comparative study for the two popular SUSY models: the
MSSM and NMSSM. We found that for both models the SUSY QCD can
leave over large residual quantum effects in $hb\bar b$ coupling
if the sparticles are relatively heavy (a few TeV) while the Higgs
sector is not so heavy ($m_A$ is not so large). Compared with the
MSSM results, the NMSSM can allow for a much broader region of
such residual effects. Since these residual effects can be over
$20\%$ in magnitude, future measurements of the $hb\bar b$
coupling may unravel such supersymmetric effects or even
distinguish the two models.

\section*{Acknowledgment}
This work was supported in part by the National Natural
Science Foundation of China under grant Nos. 10821504, 10725526 and 10635030,
and by the Natural Science Foundation of Beijing under grant No. 1072001.

\end{document}